\documentclass[12pt]{iopart}
\usepackage{iopams,bm,verbatim}
\def\ac{\overline{\alpha}}

\def\rc{\overline{\rho}}
\def\mc{\overline{\mu}}

\def\pc{\overline{\pi}}

\def\lc{\overline{\lambda}}
\def\nc{\overline{\nu}}
\def\s5{\sqrt{5}}
\def\oS{\omega_{\Sigma}}
\newcommand{\w}[1]{\bm{#1}} 

\begin{document}

\title{Rotating solenoidal perfect fluids of Petrov type D.}

\author{Norbert Van den Bergh\footnote{e-mail:
norbert.vandenbergh@ugent.be} and Lode Wylleman\footnote{e-mail:
lode.wylleman@ugent.be}}

\address{Faculty of Applied Sciences TW16, Ghent University, \\
Galglaan 2, 9000 Ghent, Belgium}

\begin{abstract}
We prove that aligned Petrov type D perfect fluids for which the vorticity vector is not orthogonal to the plane of repeated principal null directions and for which the magnetic part of the Weyl tensor with respect to the fluid velocity has vanishing divergence, are necessarily purely electric or locally rotationally symmetric. The LRS metrics are presented explicitly.
\end{abstract}

\pacs{0420}



\section{Introduction}
Recently there has been some interest in perfect fluid solutions of the Einstein field equations,
\begin{equation}
G_{ab}\equiv R_{ab}-\frac{1}{2}R g_{ab}= (w+p) u_a u_b +p g_{ab},
\end{equation}
which have been called~\cite{shearfree_solenoidal,dust} \emph{solenoidal}. These are space-times having a vanishing spatial divergence of the magnetic part of the Weyl tensor, when the latter is decomposed~\cite{Kramer} with respect to the fluid velocity $\w{u}$:
\begin{equation}\label{sol1}
(\textrm{div} \w{H})_a  = D^b H_{ab} = {h_a}^m h^{bn} H_{mb;n} = 0.
\end{equation}
The resulting models form an obvious generalization of the purely electric ($\mathbf{H}=0$) fluids, among which one finds many physically relevant exact solutions. They also offer the hope of being mathematically tractable, as one set of the Bianchi equations~\cite{Maartens1} simplifies considerably and becomes algebraic in $\w{E}, \w{\sigma}, \w{\omega}$ and $w+p$:
\begin{equation} \label{coveq}
3 E_{ab}\omega^b+(w +p) \omega_a-[\w{\sigma},\, \w{E}]_a = 0.
\end{equation}
All investigations so far seem to indicate however that the solenoidal condition ---although significantly weaker than the condition $\mathbf{H}=0$--- is still very strong and leads to classes of well known solutions, such as the locally rotationally symmetric (LRS) ones or the purely electric solutions.
In the present paper we show that if an aligned perfect fluid of Petrov type D (meaning that the fluid velocity is in the plane $\Sigma$ of the repeated principal null directions $\mathbf{k}$ and $\mathbf{l}$), is solenoidal and has a vorticity vector which is  \emph{not orthogonal} to $\Sigma$, then this fluid is necessarily purely electric or LRS. In the latter case the line-elements can be given in explicit form.

 We will make use of the Newman-Penrose
formalism and follow the notation and sign conventions
of \cite{Kramer}, whereby the Newman-Penrose equations (7.21a -- 7.21r) and Bianchi identities
(7.32a -- 7.32k) will be indicated as (np1 -- np18) and (b1 -- b11)
respectively. We choose a canonical type D tetrad, in which $\Psi_2=\phi+i \psi$ is the only non-vanishing component of the Weyl-spinor ($\phi, \psi$ real). Provided that the fluid velocity is aligned with $\Sigma$, we fix a boost such that the only non-vanishing components of the traceless Ricci spinor are given by $\Phi_{00}=\Phi_{22}=2 \Phi_{11}= S/4$, with $S=w+p$ the sum of the fluid's energy density and pressure. The energy density is then related to the Ricci scalar by the relation $R=4 w-3 S$. We also decompose the fluid's vorticity vector $\w{\omega}$ in a component $\w{\omega}_{\perp}$ orthogonal to $\Sigma$ and a component parallel to $\Sigma$. One can easily verify, see for example \cite{GoodeWainwright}, that $\w{\omega}-\w{\omega}_{\perp}$ is proportional to the following combination of the spin-coefficients, $\rc-\rho+\mc-\mu$, which we will henceforth abbreviate as
\begin{equation}
\oS = \rc-\rho+\mc-\mu
\end{equation}
and which we \emph{assume to be non-zero}.

\section{Main result}

Choosing a canonical type D tetrad as described above, the solenoidal condition (\ref{sol1}) can be expressed by the equations
\begin{eqnarray}\label{sol2a}
\delta \psi  = \frac{3}{2}(\nc-\overline{\pi}+\tau-\kappa)\psi \\
\label{sol2b} (D-\Delta)\psi  = \frac{3}{2}(\rho+\rc+\mu+\mc)\psi .
\end{eqnarray}
The Bianchi equations on the other hand lead to the differential equations
\begin{eqnarray}
\label{delS} \delta S   =  -12\kappa(\phi+i\psi)+(\kappa-\pc+2\beta+2\ac) S\\
\label{delphi} \delta \phi  =  \frac{1}{2}(2 \kappa+3\tau-3\pc) \phi + \frac{i}{2}(2 \kappa+3\tau+3\pc) \psi \nonumber\\
  \ \ \ \ -\frac{S}{24}(\kappa+\nc-\tau-\pc+4 \beta +4 \ac)  \\
\label{delpsi} \delta \psi  =  -\frac{3}{2}i(\tau+\pc)\phi+\frac{3}{2}(\tau-\pc)\psi-i \frac{S}{8}(\kappa+\nc-\tau-\pc) \\
\label{DmDelpsi} (D -\Delta)\psi  =  \frac{3}{2} (\rho+\rc+\mu+\mc) \psi +\frac{i}{4}\oS (S+6\phi)\\
(D+\Delta) \psi  = \frac{3}{2} (\rho+\rc-\mu-\mc) \psi +\frac{3}{2}i(\rc-\rho+\mu-\mc)\phi
\end{eqnarray}
and three algebraic relations, obtained by making linear combinations of (b2, b5, b7,b8), namely
\begin{eqnarray}
\label{alg1} 3(\kappa+\nc) \phi +3 i (\kappa-\nc)\psi -\frac{S}{4}(\kappa+\nc-\tau-\pc+4 \ac + 4 \beta) = 0\\
\label{sl1} 12 \sigma (\phi + i \psi)+S (\sigma-\lc) = 0\\
\label{sl2} \sigma (\phi + i \psi)+\lc (\phi-i \psi) = 0
\end{eqnarray}
Comparing (\ref{sol2b}) and (\ref{DmDelpsi}) we see that $\oS (S+6 \phi) = 0$ and hence, provided that $\oS\neq 0$,
\begin{equation}\label{Sphi}
S+6 \phi = 0.
\end{equation}
Herewith the determinant of the linear and homogeneous system (\ref{sl1}, \ref{sl2}) in $\sigma$ and $\lc$ equals $6\psi^2$. Discarding for the moment\footnote{a detailed discussion of the purely electric Petrov type D perfect fluids will be given in a forthcoming publication, see \cite{WyllemanPEtypeD}} the purely electric case (as then the condition $\textrm{div} \mathbf{H}=0$ is trivially satisfied), it follows that the null congruences corresponding to $\mathbf{k}$ and $\mathbf{l}$ are shear-free:
\begin{equation}
\sigma=\lambda = 0.
\end{equation}
In addition the equations (\ref{sol2b}), (\ref{delpsi}) and (\ref{alg1}) imply
\begin{equation}
\beta+\ac = \tau-\kappa+\pc-\nc
\end{equation}
and
\begin{equation} \label{divH}
(\kappa+\nc-3\tau-3\pc)\phi-2 i (\kappa-\nc)\psi = 0.
\end{equation}
A second algebraic relation between $\kappa,\tau,\nc$ and $\pc$ is obtained by applying the $\delta$-operator to (\ref{Sphi}): using (\ref{delS},\ref{delphi}) this yields
\begin{equation}\label{Scond}
(3\kappa-5\nc -\tau+7\pc) \phi+2 i (2\kappa-3\tau-3\pc)\psi=0.
\end{equation}
Solving (\ref{divH}) and (\ref{Scond}) for $\kappa$ and $\tau$ one finds
\begin{eqnarray}
\label{kappa} \kappa & = \left[-6\phi^2 \pc+( 2 i \phi \psi+4\phi^2-3 \psi^2)\nc \right] \left( 2 i \phi \psi+2\phi^2-3 \psi^2\right)^{-1} \\
\label{tau} \tau & = \left[ (2 i \phi \psi-4\phi^2+3 \psi^2)\pc +2(\phi^2-\psi^2)\nc \right] \left( 2 i \phi \psi+2\phi^2-3 \psi^2\right)^{-1}.
\end{eqnarray}
Next we notice that the NP-equations np2, np16 and the complex conjugates of np7, np10 give algebraic expressions for the $\delta$-derivatives of $\kappa, \tau, \nc$ and $\pc$. Substituting in these the two expressions (\ref{kappa}, \ref{tau}), two further restrictions on $\pi$ and $\nu$ follow:
\begin{eqnarray*}
\label{eq3}  \fl 3 (3 \psi^2-2 \phi^2-10 i \psi \phi) (\phi- i \psi) \phi^2 \pi^2+\phi (8 \phi^4+28 i \phi^3 \psi+38 \psi^2 \phi^2-21 \psi^4
-32 i \phi \psi^3) \nu \pi\\
-(2 \phi^5+12 i \phi^4 \psi-23 \psi^4 \phi+12 i \psi^5+21 \phi^3 \psi^2-33 i \phi^2 \psi^3) \nu^2=0\\
\label{eq2} \fl \phi (8 \phi^5+84 i \phi^4 \psi+102 \phi^3 \psi^2+54 i \psi^5+81 \psi^4 \phi+44 i \phi^2 \psi^3) \pi^2+6 i (-16 \phi^5
+26 i \phi^4 \psi\\
-12 \psi^4 \phi+18 \phi^3 \psi^2-6 i \psi^5-5 i \phi^2 \psi^3) \psi \nu \pi+(94 \psi^2 \phi^4+12 \psi^6-8 \phi^6\\
-112 i \psi^3 \phi^3+44 i \psi \phi^5+66 i \psi^5 \phi-99 \psi^4 \phi^2) \nu^2 = 0.
\end{eqnarray*}
Elimination of $\nu$ results then in
\begin{eqnarray*}
(144 \phi^8+424 \psi^2 \phi^6+848 \psi^4 \phi^4+838 \psi^6 \phi^2+297 \psi^8) (\phi+
2 i \psi)^2\\
\times (3 \psi^2-2 \phi^2+2 i \psi \phi)^4 \phi^2 \psi^2 \pi^4=0,
\end{eqnarray*}
while elimination of $\pi$ gives a similar equation for $\nu$. As the first factor is positive definite, it follows that solutions are purely electric or purely magnetic, with the second option being excluded because of (\ref{coveq}), or satisfy $\nu=\pi=0$. In the latter case $\kappa=\tau=0$ follow from (\ref{kappa},\ref{tau}) and solutions are LRS (see~\cite{GoodeWainwright}), more particularly of Ellis' class I \cite{Ellisdust, StewartEllis}.

\section{Line elements and discussion}

In this section we present the line elements of the LRS I aligned perfect fluid space-times of Petrov type D satisfying the solenoidal condition $\textrm{div} \mathbf{H}=0$ and having vorticity not orthogonal to $\Sigma$. As the general form of the rotating LRS class I perfect fluid line elements is well known~\cite{Ellisdust}, it suffices to use an algebraic computer package such as GRTensor~\cite{GRTensor} in order to explicitly express the solenoidal condition. It turns out however that the following metric form is more convenient for the actual computations:
\begin{equation}\label{metric}
ds^2=-{\w{\theta}^0}^2+{\w{\theta}^1}^2+{\w{\theta}^2}^2+{\w{\theta}^3}^2,
\end{equation}
with the duals $\w{\theta}^{i}$ of the basisvectors $\w{e}_i$ given by
\begin{eqnarray}
\w{\theta}^0 = f^{-1}s[dt+q r^2(1+\frac{k r^2}{4})^{-1}d\varphi] \nonumber \\
\w{\theta}^1 = f^{-2}dx \nonumber\\
\w{\theta}^2 = f^{-1}(1+ \frac{k r^2}{4})^{-1}dr \nonumber\\
\w{\theta}^3 =r f^{-1} (1+ \frac{k r^2}{4})^{-1} d\varphi
\end{eqnarray}
($f,s$ functions of $x$ and $k,q$ constants). This represents a perfect fluid with velocity $\mathbf{e}_0$ provided the equation of pressure isotropy holds,
\begin{equation}\label{feq}
f^2 s''-2s f f''+f s' f'-2 s {f'}^2+k s +2q^2s^3=0.
\end{equation}
The $\mathbf{e}_0$ congruence is non-expanding and shearfree, with vorticity and acceleration respectively given by
\begin{equation}
\w{\omega}=q f s\w{e}_1,\ \ \dot{\w{u}} = -s f (f/s)'\w{e}_1.
\end{equation}
The solutions therefore belong to Collins' class IIIAGii of stationary and shear-free fluids obeying a barotropic equation of state and having vorticity parallel to the acceleration~\cite{CollinsWhite}. They are of class LRS Id in the classification of Stewart and Ellis~\cite{StewartEllis} and admit a $G_4$ isometry group acting multiply transitively on the timelike hypersurfaces orthogonal to $\w{e}_1$.
Calculating the electric and magnetic part of the Weyl tensor results in
\begin{eqnarray}
\mathbf{H}=q f^3 s' \, \textrm{diag}(0,2,-1,-1),\\
\mathbf{E}= \frac{f^2}{6 s}(f^2 s''+f f' s'-4 q^2s^3-k s) \, \textrm{diag}(0,2,-1,-1).
\end{eqnarray}
while
\begin{equation}
\textrm{div}\, \mathbf{H}= 2 q f^5 s'' \w{e}_1.
\end{equation}
The solenoidal condition therefore holds when $s=1$ or $s=x$. In the first case, $s=1$, solutions are purely electric generalizations of the G\"odel metric, with $f$ given by
\begin{equation}\label{PEmetric}
f = (c_1+c_2 x +(\frac{k}{2} + q^2) x^2)^{1/2},
\end{equation}
while in the second case, $s=x$, the solution $f$ of (\ref{feq}) is given by
\begin{equation}
f=(c_1+c_2x^{3/2}+k x^2+\frac{q^2}{5}x^4)^{1/2}
\end{equation}
($c_1,c_2$ constants of integration). In the latter case the expressions for the magnetic and electric parts of the Weyl curvature reduce to
\begin{eqnarray}
\fl \mathbf{H}=q f^3 \, \textrm{diag}(0,2,-1,-1),\\
\fl \mathbf{E}= \frac{1}{200 x}(5 c_1-5 c_2 x^{3/2}+q^2 x^4+5 k x^2) (24 q^2 x^3+5 c_2 x^{1/2})\, \textrm{diag}(0,2,-1,-1).
\end{eqnarray}
Two simple examples are obtained for $c_1=c_2=0$ and $c_1=0,c_2=\frac{5 k^2}{4 q^2}$, in which the pressure $p$ and matter density $w$ are related respectively by the expressions
\begin{equation}
 p= \frac{1}{5} w -\frac{18}{25}k q^2 x^4 = \frac{3}{25} q^4 x^6
\end{equation}
and
\begin{equation}
 p= \frac{1}{5} w - \frac{9 k}{50 q^2}(5 k+2 q^2 x^2)^2 = \frac{3}{100 q^2} (q^2 x^2-5k)(2q^2 x^2+5 k)^2.
\end{equation}
In the flat case, $k=0$, both metrics reduce to
\begin{equation}\label{NLmetric}
ds^2=5q^{-2}x^{-2}\left( -(dt+q r^2 d\phi)^2+5q^{-2}x^{-6} dx^2+x^{-2}(dr^2+r^2 d\phi^2)\right).
\end{equation}
Introducing standard coordinates by putting $x=\sqrt{5}/(q z)$ this can also be written as
\begin{equation}\label{Wform}
ds^2 = - z^2 (dt + q r^2 d \varphi)^2 +\frac{q^2 z^4}{5} ds_3^2,
\end{equation}
with $ds_3^2=dr^2+r^2 d\varphi^2+dz^2$ the metric of 3-d Euclidean space. The solution clearly satisfies then the regularity condition on the $r=0$ axis and represents a rigidly rotating perfect fluid obeying a $\gamma$-law equation of state $p=\frac{1}{5} w$. This particular equation of state suggests a possible link with the Collins-Stewart~\cite{CollinsStewart} purely magnetic (spatially homogeneous and expanding, but non-rotating) perfect fluid metric
\begin{equation}\label{CSmetric}
ds^2= -\alpha^2 dt^2+t^{2/3}(dx+r^2d\phi)^2+t^{4/3}(dr^2+r^2d\phi^2),\ \ \alpha^2=\frac{1}{3}.
\end{equation}
However, although a complex coordinate transformation $t\rightarrow q x,\, x\rightarrow i q^{-1}t^{-1/3}$ transforms (\ref{NLmetric}) into a metric of the form (\ref{CSmetric}) (modulo a global scale factor), the resulting constant $\alpha^2$ does not have the correct value $\alpha^2=\frac{1}{3}$.
The metric (\ref{Wform}) belongs to a class of self-similar solutions with an $H_5$ similarity group discussed by Wainwright in \cite{Bonnorvolume}.

\section*{References}

\providecommand{\newblock}{}

\end{document}